\documentclass[english]{article}
\usepackage[T1]{fontenc}
\usepackage[latin9]{inputenc}
\usepackage{geometry}
\usepackage{amsmath}
\usepackage{amsthm}
\usepackage{amssymb}
\usepackage{setspace}
\makeatletter
\theoremstyle{plain}
\newtheorem{thm}{\protect\theoremname}
\theoremstyle{definition}
\newtheorem{defn}[thm]{\protect\definitionname}
\theoremstyle{plain}
\newtheorem{fact}[thm]{\protect\factname}
\theoremstyle{plain}
\newtheorem{cor}[thm]{\protect\corollaryname}
\theoremstyle{definition}
\newtheorem{example}[thm]{\protect\examplename}
\theoremstyle{plain}
\newtheorem{lem}[thm]{\protect\lemmaname}
\newenvironment{lyxlist}[1]
	{\begin{list}{}
		{\settowidth{\labelwidth}{#1}
		 \setlength{\leftmargin}{\labelwidth}
		 \addtolength{\leftmargin}{\labelsep}
		 }}
	{\end{list}}

\usepackage{fullpage}

\usepackage{graphicx}
\usepackage{xcolor}
\usepackage{transparent}

\usepackage{authblk}

\usepackage{microtype}

\usepackage{todonotes}
\usepackage{textpos}

\usepackage[all=normal,bibliography=tight]{savetrees}

\makeatother

\usepackage{babel}
\providecommand{\corollaryname}{Corollary}
\providecommand{\definitionname}{Definition}
\providecommand{\examplename}{Example}
\providecommand{\factname}{Fact}
\providecommand{\lemmaname}{Lemma}
\providecommand{\theoremname}{Theorem}

\begin{document}
\title{Count on CFI graphs for \#P-hardness\thanks{\rightskip=5cm 
Funded by the European Union (ERC, CountHom, 101077083). Views and opinions expressed are however those of the author(s) only and do not necessarily reflect those of the European Union or the European Research Council Executive Agency. Neither the European Union nor the granting authority can be held responsible for them. The author is also part of BARC, supported by the VILLUM Foundation grant 16582.}
\author{Radu Curticapean}
\affil{IT University of Copenhagen, Denmark}
\begin{textblock}{5}(8, 10.5) \includegraphics[width=100px]{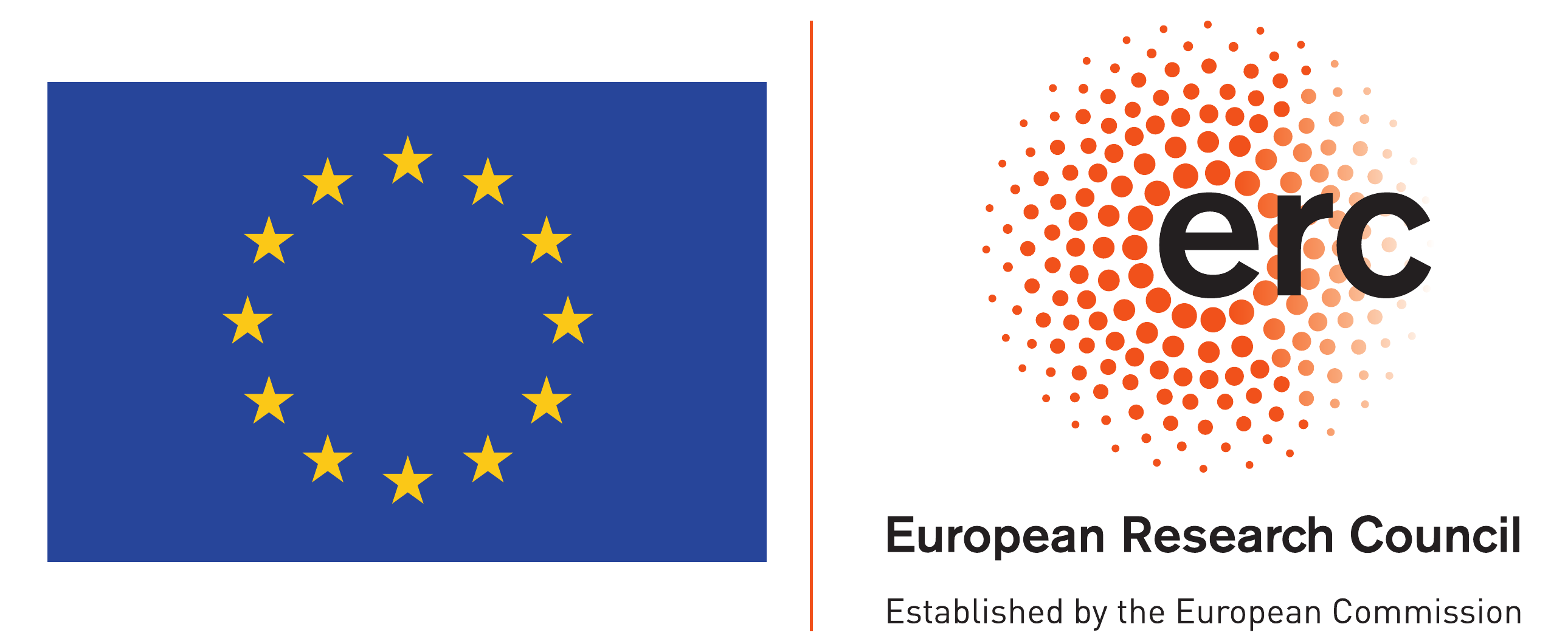} \end{textblock}
\date{}}
\maketitle
\begin{abstract}
\global\long\def\GF{\mathrm{GF}}%
\global\long\def\ext{\mathrm{Ext}}%
\global\long\def\quot{\mathrm{Quot}}%
\global\long\def\inhom{\triangleleft\,}%
\global\long\def\tw{\mathrm{tw}}%
\global\long\def\surj{\mathrm{surj}}%
\global\long\def\wall{\mathrm{wall}}%
\global\long\def\htw{\mathrm{htw}}%
\global\long\def\ind{\mathrm{ind}}%
\global\long\def\emb{\mathrm{emb}}%
\global\long\def\sub{\mathrm{sub}}%
\global\long\def\pcdot{\,\cdot\,}%
\global\long\def\ssneq{\sqsubsetneq}%
\global\long\def\aut{\mathrm{aut}}%
\global\long\def\SAT{\mathrm{\#SAT}}%
\global\long\def\Hom{\mathrm{\#Hom}}%
\global\long\def\PartSub{\mathrm{\#PartitionedSub}}%
\global\long\def\ColHom{\mathrm{\#ColHom}}%
\global\long\def\Sub{\mathrm{\#Sub}}%
\global\long\def\Ind{\mathrm{\#Ind}}%
\global\long\def\FP{\mathsf{FP}}%
\global\long\def\NP{\mathsf{NP}}%
\global\long\def\P{\mathsf{P}}%
\global\long\def\sharpP{\mathsf{\#P}}%
\global\long\def\FPT{\mathsf{FPT}}%
\global\long\def\sharpWone{\mathsf{\#W[1]}}%
\global\long\def\ETH{\mathsf{ETH}}%
\global\long\def\sharpETH{\mathsf{\#ETH}}%
\global\long\def\delcols#1{\negthinspace\downharpoonright_{#1}}%

Given graphs $H$ and $G$, possibly with vertex-colors, a \emph{homomorphism}
is a function $f:V(H)\to V(G)$ that preserves colors and edges. Many
interesting counting problems are finite linear combinations $p(\pcdot)=\sum_{H}\alpha_{H}\hom(H,\pcdot)$
of homomorphism counts from fixed graphs $H$. It is known that such
fixed linear combinations are hard to evaluate under standard complexity
assumptions if there exists a large-treewidth graph $S$ with $\alpha_{S}\neq0$.
This can be shown in two steps: In the first step, hardness is shown
for the problems $\hom(S,\pcdot)$ with colorful (i.e., bijectively
colored) graphs $S$ of large treewidth. In the second step, it is
shown that homomorphism counts from colorful graphs $S$ into graphs
$G$ can be reduced to evaluating finite linear combinations of homomorphism
counts that contain the uncolored version $S^{\circ}$ of $S$. This
second step can be performed via inclusion-exclusion in $2^{|E(S)|}\mathrm{poly}(n,s)$
time, where $n$ is the size of $G$ and $s$ is the maximum number
of vertices among graphs in the linear combination. This running time
is acceptable when $S$ and $s$ are small compared to $n$, which
is the setting considered in previous works.

In this paper, we show that the second step can even be performed
in time $4^{\Delta(S)}\mathrm{poly}(n,s)$, where $\Delta(S)$ is
the maximum degree of $S$. Our reduction is based on graph products
with Cai--Fürer--Immerman graphs, a novel technique that is likely
of independent interest. If $\Delta(S)$ is constant and $s$ is polynomial
in $n$, this technique yields a polynomial-time reduction from $\hom(S,\pcdot)$
to linear combinations of homomorphism counts involving $S^{\circ}$,
even when $S$ is part of the input. Under certain conditions, it
actually suffices that a \emph{supergraph} $T$ of $S^{\circ}$ is
contained in the target linear combination, which is useful, since
large treewidth is witnessed by subgraphs of maximum degree $3$.

Using the new reduction, we obtain $\sharpP$-hardness results for
problems that could previously only be studied under parameterized
complexity assumptions that are a priori stronger than ``classical''
assumptions. For fixed graph classes $\mathcal{H}$ satisfying natural
polynomial-time enumerability conditions, this includes the problems
$\Hom(\mathcal{H})$, $\Sub(\mathcal{H})$ and $\Ind(\mathcal{H})$
of counting, on input a graph $H\in\mathcal{H}$ and a general graph
$G$, the homomorphisms from $H$ to $G$, and the subgraph copies
and induced subgraph copies of $H$ in $G$. Generally speaking, the
new reduction opens up the possibility of transferring techniques
from parameterized counting complexity to ``classical'' non-parameterized
counting complexity.
\end{abstract}
\clearpage{}

\section{Introduction}

In this paper, graphs $G$ are undirected and may be vertex-colored,
i.e., they may be given with a coloring $c:V(G)\to C$ for some set
$C$ that is specified from the context. We write $G\simeq G'$ if
two graphs $G$ and $G'$ admit a color-preserving isomorphism. Uncolored
graphs can be viewed as colored graphs with a single color.

A graph parameter is a function $p$ that maps graphs into $\mathbb{Q}$
such that $p(G)=p(G')$ holds for isomorphic graphs $G\simeq G'$.
Graph parameters form a vector space under scalar multiplication and
point-wise addition, with canonical basis functions given by the graph
parameters $e_{H}$ for fixed unlabeled graphs $H$, which satisfy
$e_{H}(G)=1$ for $G\simeq H$ and $e_{H}(G)=0$ otherwise. Note that
only $G$ is part of the input for $e_{H}$. Other examples for graph
parameters include the functions that map a graph $G$ to its maximum
degree, maximum clique size, number of perfect matchings, or the evaluation
of the Tutte polynomial $T(G;x,y)$ for fixed $x,y\in\mathbb{Q}$,
which is a weighted sum over subgraphs of $G$. In this paper, we
focus on graph parameters that admit obvious interpretations as weighted
counts of objects in graphs, such as the last two examples.

\subsection{Graph motif parameters}

Many interesting graph parameters that count objects are \emph{homomorphism
counts} or finite linear combinations thereof. A homomorphism from
a graph $H$ to a graph $G$ is a function $f:V(H)\to V(G)$ such
that $uv\in E(H)$ implies $f(u)f(v)\in E(G)$. If $H$ and $G$ are
colored, then the colors of $v$ and $f(v)$ must agree for all $v\in V(H)$.
For any fixed graph $H$, let $\hom(H,\pcdot)$ be the graph parameter
that maps graphs $G$ to the number $\hom(H,G)$ of homomorphisms
from $H$ to $G$. As in the definition of the functions $e_{H}$
above, only $G$ is part of the input.

The functions $\hom(H,\pcdot)$ are known to be linearly independent~\cite[Proposition~5.43]{Lovasz2012}:
If the all-zeros graph parameter is expressed as a finite linear combination
of these functions, for pairwise non-isomorphic graphs, then this
linear combination must be trivial. It follows that no two distinct
finite linear combinations of the functions $\hom(H,\pcdot)$ yield
the same graph parameter. This leads to the following definitions:
\begin{defn}[see \cite{DBLP:conf/stoc/CurticapeanDM17}]
\label{def: hom-expansion}A \emph{graph motif parameter} $p$ is
a graph parameter 
\begin{equation}
p(\pcdot)=\sum_{H\in\mathcal{H}}\alpha_{H}\hom(H,\pcdot)\label{eq: gen-graph-motif}
\end{equation}
for a finite set of pairwise non-isomorphic graphs $\mathcal{H}$
and coefficients $\alpha_{H}\in\mathbb{Q}$ for $H\in\mathcal{H}$.
We call the right-hand side of (\ref{eq: gen-graph-motif}) the \emph{hom-expansion}
of $p$. Given a graph $H$, we write $H\inhom p$ if $H\in\mathcal{H}$
and $\alpha_{H}\neq0$.
\end{defn}

Besides the functions $\hom(H,\pcdot)$ for fixed graphs $H$, other
examples for graph motif parameters are the functions $\sub(H,\pcdot)$
that count subgraphs isomorphic to fixed graphs $H$:
\begin{fact}
\textup{\label{fact: sub hom-exp}For any graph $H$, the function
}$\sub(H,\pcdot)$\textup{ is a graph motif parameter, see~\cite[(5.18)]{Lovasz2012}
. We have $F\inhom\sub(H,\pcdot)$ iff $F$ is a }\textup{\emph{quotient}}\textup{
of $H$, i.e., if $F$ can be obtained from $H$ by repeated identifications
of vertices of the same color. The sign of the coefficient $\alpha_{F}$
in the hom-expansion of $\sub(H,\pcdot)$ equals $(-1)^{|V(H)|-|V(F)|}$.}
\end{fact}

A similar statement holds for the functions $\ind(H,\pcdot)$ that
count induced subgraphs isomorphic to $H$.
\begin{fact}
\textup{\label{fact: ind hom-exp}For any graph $H$, the function
$\ind(H,\pcdot)$ is a graph motif parameter, see~\cite[(5.20)]{Lovasz2012}.
Moreover, any graph $F\inhom\ind(H,\pcdot)$ satisfies $|V(F)|\leq|V(H)|$.
The graphs $F$ with $F\inhom\ind(H,\pcdot)$ and $|V(F)|=|V(H)|$
are precisely the graphs obtained by adding edges to $H$.}
\end{fact}

Many counting problems related to small patterns can be expressed
naturally as linear combinations of (induced) subgraph counts, and
Facts~\ref{fact: sub hom-exp} and \ref{fact: ind hom-exp} then
imply that they are graph motif parameters. Examples include the numbers
of vertex-subsets inducing a graph from some fixed finite set of graphs
$\mathcal{P}$~\cite{DBLP:conf/focs/Roth0W20,DBLP:conf/iwpec/RothS18,Jerrum2015,Jerrum2015a,Jerrum2016},
or variants of the Tutte polynomial that only sum over subgraphs with
a fixed number of edges~\cite{DBLP:conf/icalp/Roth0W21}.

.

\subsection{Known complexity monotonicity}

Any fixed graph motif parameter $p$ can be evaluated in polynomial
time on $n$-vertex input graphs: First, we can evaluate $\hom(H,G)$
for all graphs $H\inhom p$ by brute-force in time $O(n^{s})$, where
$s$ is the maximum of $|V(H)|$ among graphs $H\inhom p$, and then
we can compute the required finite linear combination with a finite
number of arithmetic operations.\footnote{By a similar argument, the actual output values of graph motif parameters
are bounded by $O(n^{s})$. This implies, e.g., that the counts of
perfect matchings or Hamiltonian cycles in graphs are not graph motif
parameters.} Departing from the polynomial-time solvability, one may then wish
to pinpoint the precise exponent in the polynomial running time or
investigate a more ``coarse-grained'' question: Given an infinite
family of graph motif parameters $\{p_{i}\}_{i\in I}$ for some index
set $I$, can we evaluate $p_{i}(G)$ in polynomial time on input
$i\in I$ and $G$? This question captures well-studied counting problems
like $\Sub(\mathcal{H})$ for fixed graph classes $\mathcal{H}$,
which asks to count occurrences of an input graph $H\in\mathcal{H}$
in general graphs $G$.

Previous works developed an approach for investigating the complexity
of graph motif parameters through their hom-expansions~\cite{DBLP:conf/stoc/CurticapeanDM17}.
This approach has been dubbed the \emph{complexity monotonicity framework}
and was applied successfully to various counting problems~\cite{DBLP:conf/esa/Roth17,DBLP:conf/focs/Roth0W20,DBLP:conf/icalp/Roth0W21,DBLP:conf/iwpec/RothS18,DBLP:conf/mfcs/DorflerRSW19,DBLP:conf/mfcs/PeyerimhoffR0SV21,DBLP:conf/stoc/CurticapeanDM17,DBLP:conf/focs/0002R21}.
The framework leverages the remarkable fact that evaluating any fixed
linear combination of homomorphism counts is asymptotically as hard
as evaluating its hardest terms. The upper bound in this statement
is trivial, while the lower bound follows from a reduction which computes
$\hom(S,G)$ for graphs $S\inhom p$ and $G$ with access to an oracle
for $p(\pcdot)$. The running time of this reduction is polynomial
in $|V(G)|$ and in the maximum size of graphs $H\inhom p$, but exponential
in the size of $S$.
\begin{thm}[variants shown in \cite{DBLP:conf/stoc/CurticapeanDM17,Chen2016}]
\label{thm: hom-monoton}There is an algorithm for the following
problem: Output the value $\hom(S,G)$ when given 
\begin{itemize}
\item as input a number $s\in\mathbb{N}$, graphs $S$ and $G$, and
\item oracle access for a graph motif parameter $p$ with $\max_{H\inhom p}|V(H)|\leq s$
and $S\inhom p$.
\end{itemize}
The running time of the algorithm is bounded by $2^{|E(S)|}\cdot\mathrm{poly}(|V(G)|,s)$,
and the oracle is only called on graphs with $\leq s\cdot|V(G)|$
vertices.
\end{thm}

Let us note that other natural bases of graph motif parameters do
not share such favorable properties. For example, counting $k$-cliques
is ``hard''~\cite{Marx2010,Chen2006a}, but the sum of induced
subgraph counts of all $k$-vertex graphs can be evaluated trivially.

\subsubsection*{Implications for individual graph motif parameters}

Theorem~\ref{thm: hom-monoton} reduces complexity-theoretic questions
about graph motif parameters $p$ to questions about homomorphism
counts $\hom(H,\pcdot)$ for graphs $H\inhom p$. These problems are
quite well-understood in various computational models~\cite{DBLP:conf/icalp/Komarath0R22}
and under complexity-theoretic assumptions such as the counting exponential-time
hypothesis $\sharpETH$~\cite{Impagliazzo2001a,Dell2014}, which
postulates that counting satisfying assignments to Boolean formulas
on $n$ variables requires $\exp(\Omega(n))$ time. Under this hypothesis,
it is known that the complexity of counting homomorphisms is essentially
determined by the \emph{treewidth} $\tw(H)$, a graph width measure
that is polynomially related to the maximum $r\in\mathbb{N}$ such
that $H$ contains an $r\times r$ grid minor~\cite{Cygan2015,Chekuri2016}.
\begin{thm}[\cite{Dalmau2004,Marx2010,DBLP:conf/stoc/CurticapeanDM17}]
\label{thm: lower-bd}For any graph $H$ of treewidth $t$, the graph
parameter $\hom(H,\pcdot)$ can be evaluated in time $O(n^{t+1})$
on $n$-vertex graphs. Assuming $\sharpETH$, there exists a constant
$c>0$ such that, for any graph $H$ of treewidth $t$, the graph
parameter $\hom(H,\pcdot)$ cannot be evaluated in time $O(n^{ct/\log t})$.
\end{thm}

Together with Theorem~\ref{thm: hom-monoton}, we obtain the following
corollary.
\begin{cor}[\cite{DBLP:conf/stoc/CurticapeanDM17}]
\label{cor: graphmotif}Let $p$ be a graph motif parameter and let
$q=\max_{H\inhom p}\tw(H)$. Then $p$ can be evaluated in time $O(n^{q+1})$.
Assuming $\sharpETH$, there is a constant $c>0$ that is independent
of $p$, such that $p$ cannot be evaluated in time $O(n^{cq/\log q})$.
\end{cor}

This corollary allows us to study the complexity of a graph motif
parameter $p$ by determining the maximum treewidth of graphs $H\inhom p$
in the hom-expansion of $p$.
\begin{example}
\label{exa: sub}Following~\cite{DBLP:conf/stoc/CurticapeanDM17},
let us investigate the complexity of $\sub(H,\pcdot)$ for fixed $H$.
By Fact~\ref{fact: sub hom-exp}, we have $F\inhom\sub(H,\pcdot)$
iff $F$ is a quotient of $H$, so it suffices to determine the maximum
treewidth $q$ that can be attained by quotients of $H$. Up to constant
factors, $q$ equals the maximum size $k$ of a matching in $H$:
\begin{itemize}
\item Every graph $Q$ on $k$ edges is a quotient of a matching on $k$
edges. Thus, for every graph $Q$ on $k$ edges, some quotient of
$H$ contains $Q$ as a subgraph. As we can choose $Q$ to be an expander
with $k$ edges and treewidth $\Omega(k)$, it follows that $q\geq\Omega(k)$.
\item Since $H$ has no matching of size $k+1$, it has a vertex-cover of
size $\leq2k$. Because taking quotients does not increase the vertex-cover
number of a graph, any quotient of $H$ also has a vertex-cover of
size $\leq2k$. It follows that $q\leq O(k)$, since the treewidth
of any graph is upper-bounded by its vertex-cover number.
\end{itemize}
Together, these two facts imply that $q=\Theta(k)$. It follows by
Corollary~\ref{cor: graphmotif} that $\sub(H,\pcdot)$ can be evaluated
in time $n^{O(k)}$ and, assuming $\sharpETH$, not in time $n^{o(k/\log k)}$.
\end{example}

It can be shown similarly that $\ind(H,\pcdot)$ with $|V(H)|=k$
cannot be solved in time $n^{o(k)}$ under $\sharpETH$: We use that
the complete graph $K_{k}$ satisfies $K_{k}\inhom\ind(H,\pcdot)$
by Fact~\ref{fact: ind hom-exp} and that $\hom(K_{k},\pcdot)$ cannot
be solved in time $n^{o(k)}$ under $\sharpETH$, see~\cite{Chen2006a}.
Similar results are known for other graph motif parameters related
to restricted types of homomorphisms~\cite{DBLP:conf/esa/Roth17},
vertex-induced subgraphs inducing monotone properties~\cite{DBLP:conf/focs/Roth0W20,DBLP:conf/iwpec/RothS18,DBLP:conf/mfcs/DorflerRSW19},
edge-induced subgraphs satisfying minor-closed properties~\cite{DBLP:conf/mfcs/PeyerimhoffR0SV21,DBLP:conf/icalp/Roth0W21},
and a variant of the Tutte polynomial~\cite{DBLP:conf/icalp/Roth0W21}.

\subsubsection*{Implications for families of graph motif parameters}

Rather than focusing on \emph{individual} graph motif parameters,
one can also study \emph{families} of graph motif parameters~\cite{Dalmau2004,Chen2008,DBLP:conf/focs/CurticapeanM14},
as is the focus of this paper. In previous works, such problems were
addressed in the framework of \emph{parameterized complexity theory}~\cite{Cygan2015,Flum2006,Downey1999},
where inputs to computational problems come in the form $(x,k)$ with
a traditional input $x$ and a \emph{parameter value} $k$ that measures
some notion of complexity in $x$. Often, the parameter is actually
the value of a graph parameter, e.g., the treewidth. Parameterized
problems can usually be solved in time $n^{g(k)}$ for some function
$g$, and they are \emph{fixed-parameter tractable} if they can even
be solved in time $f(k)n^{O(1)}$ for some computable function $f$.

The evaluation of graph motif parameters naturally leads to a parameterized
problem~\cite{DBLP:conf/stoc/CurticapeanDM17}: Given as input coefficients
$\alpha_{1},\ldots,\alpha_{t}\in\mathbb{Q}$, graphs $H_{1},\ldots,H_{t}$,
and an additional graph $G$, the task is to evaluate $p(G)=\sum_{i=1}^{t}\alpha_{i}\hom(H_{i},G)$.
The parameter in this problem is $\max_{i}|V(H_{i})|$. This general
problem subsumes the problem of counting $k$-cliques in graphs, the
canonical hard problem for the complexity class $\sharpWone$, which
is the parameterized analogue of $\NP$. More generally, a parameterized
problem is $\sharpWone$-hard if counting $k$-cliques can be reduced
to it via a version of Turing reductions that is adapted to parameterized
problems.

The above evaluation problem for graph motif parameters can be studied
in a more refined way by fixing a recursively enumerable family $\mathcal{A}$
of graph motif parameters (represented as lists of coefficients and
graphs) and considering the evaluation problem under the promise that
$p\in\mathcal{A}$. One specific example is the problem $\Hom(\mathcal{H})$
for recursively enumerable graph classes $\mathcal{H}$: On input
$H\in\mathcal{H}$ and $G$, determine $\hom(H,G)$, parameterized
by $|V(H)|$. This problem is known to be polynomial-time solvable
if the maximum treewidth among graphs in $\mathcal{H}$ can be bounded
by a constant, while it is $\sharpWone$-hard otherwise~\cite{Dalmau2004}.
Together with Theorem~\ref{thm: hom-monoton}, this implies the following
classification of the evaluation problem for graph motif parameters:
\begin{thm}[\cite{DBLP:conf/stoc/CurticapeanDM17}]
\label{thm: main-thm-old}Let $\mathcal{A}$ be a recursively enumerable
family of graph motif parameters. If the maximum treewidth of graphs
$F\inhom p$ for $p\in\mathcal{A}$ is unbounded, then evaluating
$p(G)$ on input $p\in\mathcal{A}$ and $G$ is $\sharpWone$-hard.
Otherwise, the problem is fixed-parameter tractable.
\end{thm}

For example, consider the problem $\Sub(\mathcal{H})$ for fixed recursively
enumerable graph classes $\mathcal{H}$, defined analogous to $\Hom(\mathcal{H})$.
By combining Theorem~\ref{thm: main-thm-old} and Fact~\ref{fact: sub hom-exp},
this problem is $\sharpWone$-hard if the vertex-cover number of graphs
in $\mathcal{H}$ is unbounded. Otherwise, the problem is fixed-parameter
tractable, and actually even polynomial-time solvable~\cite{Williams2013,DBLP:conf/focs/CurticapeanM14}.
Likewise, the problem $\Ind(\mathcal{H})$ is $\sharpWone$-hard if
$\mathcal{H}$ is infinite and polynomial-time solvable otherwise.
While classifications for these problems were known before the introduction
of the complexity monotonicity framework~\cite{Dalmau2004,Chen2008,DBLP:conf/focs/CurticapeanM14},
that framework enabled significantly simpler proofs that abstracted
away most of the specifics of the problems and enabled the study of
further problems~\cite{DBLP:conf/esa/Roth17,DBLP:conf/focs/Roth0W20,DBLP:conf/icalp/Roth0W21,DBLP:conf/iwpec/RothS18,DBLP:conf/mfcs/DorflerRSW19,DBLP:conf/mfcs/PeyerimhoffR0SV21,DBLP:conf/stoc/CurticapeanDM17}.

\subsection{\label{subsec: comp-mon-polytime}Complexity monotonicity via polynomial-time
reductions}

The main technical result of the present paper is a \emph{polynomial-time
variant} of Theorem~\ref{thm: hom-monoton}, the central reduction
of the complexity monotonicity framework. As in Theorem~\ref{thm: hom-monoton},
we obtain a Turing reduction that computes $\hom(S,\pcdot)$ for a
graph $S$ with an oracle for a graph motif parameter $p$ with $S\inhom p$,
but our new reduction runs in polynomial time when $S$ has bounded
degree. Additionally, the new reduction allows us to reduce homomorphism
counts of \emph{colorful} graphs to the evaluation of an \emph{uncolored}
graph parameter. This is very useful, as $\sharpP$-hardness of counting
colorful homomorphisms can be shown relatively straightforwardly,
while the required $\sharpP$-hardness proofs for uncolored graphs
were not known prior to this paper.

In the following, we call a colored graph $H$ \emph{colorful} if
its coloring is a bijection. Unless stated otherwise, we assume that
the color set of a colorful graph $H$ is $V(H)$ and that the coloring
is the identity function. We also write $H^{\circ}$ for the uncolored
graph underlying $H$, and we note that $\hom(H,G)\neq\hom(H^{\circ},G^{\circ})$
in general.
\begin{thm}
\label{thm: poly-mon}There is an algorithm for the following problem:
Output the value $\hom(S,G)$ when given 
\begin{itemize}
\item as input a number $s\in\mathbb{N}$, colorful graphs $S$ and $T$
with $S\subseteq T$, a colored graph $G$, and
\item oracle access for a graph motif parameter $p$ with $\max_{H\inhom p}|V(H)|\leq s$
such that $T^{\circ}\inhom p$ and either
\begin{enumerate}
\item[(a)] $S=T$, or 
\item[(b)] all graphs on $|V(T)|$ vertices in the hom-expansion of $p$ have
the same sign, or
\item[(c)] no proper edge-subgraph of $T^{\circ}$ is contained in the hom-expansion
of $p$.
\end{enumerate}
\end{itemize}
The running time of the algorithm is bounded by $4^{\Delta(S)}\cdot\mathrm{poly}(|V(G)|,s)$.
\end{thm}

Let us elaborate on the differences between the known reduction and
our new variant. Most importantly, the exponential dependence in the
running time stated in Theorem~\ref{thm: poly-mon} is confined to
the maximum degree $\Delta(S)$ rather than $|E(S)|$. In particular,
the reduction runs in polynomial time if $S$ has bounded maximum
degree, even when the sizes of $S$ and $G$ are comparable. Note
also that we reduce from $S$ but only require a supergraph $T\supseteq S$
with $T^{\circ}\inhom p$, subject to one of the three technical conditions
of the theorem. This is very useful, since any graph $T$ of large
treewidth contains a large \emph{wall}, which is a subgraph $S\subseteq T$
of large treewidth and maximum degree $3$. We can then reduce from
$S$ and consider the remainder of $T$ as ``dummy'' vertices and
edges, thus avoiding an exponential running time even if $\Delta(T)$
is large.

Maybe more subtly, note that we reduce $\hom(S,\cdot)$ for colorful
graphs $S$ to evaluating an uncolored graph parameter $p$. This
is already interesting when $p=\hom(S^{\circ},\cdot)$ counts uncolored
homomorphisms from $S^{\circ}$, as the previously known reductions
require $2^{|E(S)|}n^{O(1)}$ time. Because of this, comprehensive
classifications for $\Hom(\mathcal{H})$ in terms of polynomial-time
and $\sharpP$-hard classes $\mathcal{H}$ were not attainable with
previous techniques. For comparison, it is somewhat easier to reduce
$\hom(S,\cdot)$ to $\hom(R,\cdot)$ for an uncolored graph $R$ derived
from $S$, say, by adding gadgets to $S$ that identify colors~\cite{DBLP:conf/icalp/BringmannS21}.
In the same spirit, there are elementary hardness proofs for deciding
the existence of colorful $k+k$ bicliques and $k\times k$ grids
in colored graphs, but the known hardness results for the \emph{uncolored}
version of these problems require significant technical effort~\cite{DBLP:conf/wg/ChenGL17,Lin2015}. 

\subsubsection*{Implications for families of graph motif parameters}

Our new Theorem~\ref{thm: poly-mon} implies $\sharpP$-hardness
results for several important families of graph motif parameters.
This shows that techniques devleoped for the complexity monotonicity
framework can also yield polynomial-time reductions and $\sharpP$-hardness
results instead of $\sharpWone$-hardness results.

A first application concerns the problems $\Hom(\mathcal{H})$ mentioned
above. The original complexity classification for these problems is
a highly cited result in parameterized complexity and database theory,
and it showed $\sharpWone$-hardness for recursively enumerable graph
classes $\mathcal{H}$ of unbounded treewidth~\cite{Dalmau2004}.
Together with a simple \emph{polynomial-time} (rather than merely
fixed-parameter tractable) algorithm for graph classes of bounded
treewidth, this classifies the \emph{polynomial-time} solvable cases
of $\Hom(\mathcal{H})$ under the assumption $\FPT\neq\sharpWone$,
subject to the condition that $\mathcal{H}$ is recursively enumerable.
Our new theorem imposes a natural polynomial-time enumerability condition
on $\mathcal{H}$ and classifies the polynomial-time solvable cases
of $\Hom(\mathcal{H})$ under the weaker assumption $\FP\neq\sharpP$.
To state the criterion, let us define an $r\times r$ \emph{wall }to
be any graph obtained from the $r\times r$ square grid by deleting
even-indexed edges from even-indexed columns and odd-indexed edges
from odd-indexed columns, and then subdividing edges arbitrarily often.
Any graph of treewidth $t$ contains an $r\times r$ wall subgraph
$S$ with $r\in\Omega(t^{1/100})$, which can be found with a randomized\footnote{Randomization does not seem to be inherently required to find the
wall subgraph, but no deterministic polynomial-time algorithm is known
to the best of our knowledge. Consequently, some of our $\sharpP$-completeness
results require randomized rather than deterministic reductions.} polynomial-time algorithm~\cite{Chekuri2016}. Our enumerability
condition is based on walls.
\begin{thm}
\label{thm: hard-hom}If a graph class $\mathcal{H}$ admits a polynomial-time
algorithm that outputs a graph $H\in\mathcal{H}$ of treewidth $\geq t$
on input $t\in\mathbb{N}$ in unary, then $\Hom(\mathcal{H})$ is
$\sharpP$-complete under randomized polynomial-time reductions. 

If the algorithm even outputs $H$ together with a $t\times t$ wall
subgraph, then $\Hom(\mathcal{H})$ is $\sharpP$-complete under deterministic
polynomial-time reductions.
\end{thm}

The theorem is shown by combining our new Theorem~\ref{thm: poly-mon}
with a standard $\sharpP$-hardness proof for counting homomorphisms
from colorful walls. Note that some kind of polynomial-time constructibility
condition is required in view of ``pathologically padded'' classes
like $\mathcal{P}=\{K_{k}\,\cup\,\overline{K_{2^{k}}}\mid k\in\mathbb{N}\}$:
Any polynomial-time algorithm for $\Hom(\mathcal{P})$ would render
the $k$-clique problem fixed-parameter tractable and thus imply $\FPT=\sharpWone$,
which in turn refutes $\sharpETH$. At the same time, the problem
$\Hom(\mathcal{P})$ is unlikely to be $\sharpP$-hard: Since it can
be solved in quasi-polynomial time, its $\sharpP$-hardness would
refute $\sharpETH$ among other complexity-theoretic assumptions.
Note that $\mathcal{P}$ does not satisfy the conditions of the theorem.

A classification similar to Theorem~\ref{thm: hard-hom} can be shown
for counting \emph{subgraphs}. The original complexity classification~\cite{DBLP:conf/focs/CurticapeanM14,DBLP:conf/stoc/CurticapeanDM17}
established $\sharpWone$-hardness for recursively enumerable graph
classes $\mathcal{H}$ of unbounded vertex-cover. As discussed in
Example~\ref{exa: sub}, if $H$ has a matching on $t$ edges, then
every graph $S$ on $t$ edges is contained as a subgraph in some
quotient $T$ of $H$, and that quotient $T$ and its subgraph $S$
can be found efficiently. By Fact~\ref{fact: sub hom-exp}, we have
$T\inhom p$. In particular, we can let $S$ be the colorful version
of a large wall without subdivisions. We obtain the following natural
analogue in terms of $\sharpP$-hardness:
\begin{thm}
\label{thm: hard-sub}If a graph class $\mathcal{H}$ admits a polynomial-time
algorithm that outputs a graph $H\in\mathcal{H}$ of vertex-cover
number $\geq t$ on input $t\in\mathbb{N}$ in unary, then $\Sub(\mathcal{H})$
is $\sharpP$-complete.
\end{thm}

The proof of this theorem also starts from the $\sharpP$-hardness
of counting colorful walls, and it invokes Theorem~\ref{thm: poly-mon}
to reduce this problem to $\Sub(\mathcal{H})$. As opposed to Theorem~\ref{thm: hard-hom},
the relevant walls in the quotients of $H$ can be found \emph{deterministically}
in polynomial time, because the process described in Example~\ref{exa: sub}
lends itself to a deterministic polynomial-time algorithm.

Finally, by a proof similar to those for Theorems~\ref{thm: hard-hom}
and \ref{thm: hard-sub}, we can also classify induced subgraph counting
problems. For these problems, we rely on polynomial-time algorithms
for finding large clique minors in graphs of large average degree
to find large walls in $H\in\mathcal{H}$ or the complement of $H$.
\begin{thm}
\label{thm: hard-ind}If a graph class $\mathcal{H}$ admits a polynomial-time
algorithm that outputs a graph $H\in\mathcal{H}$ on $\geq t$ vertices
on input $t\in\mathbb{N}$ in unary, then $\Ind(\mathcal{H})$ is
$\sharpP$-complete.
\end{thm}

Let us remark that the main technical effort lies in proving Theorem~\ref{thm: poly-mon},
while Theorems~\ref{thm: hard-hom}--\ref{thm: hard-ind} follow
fairly straightforwardly using known techniques. We remark that the
$\sharpP$-hardness results from Theorems~\ref{thm: hard-hom}--\ref{thm: hard-ind}
are complemented by straightforward polynomial-time algorithms when
the relevant graph parameters (treewidth, vertex-cover number, size)
are bounded in a class $\mathcal{H}$.

\subsection{\label{subsec: proof-ideas}Proof outline}

The proof of Theorem~\ref{thm: poly-mon} is built around several
\emph{filters} that allow us to progressively narrow the hom-expansion
of $p$ down to the desired graph $S$. In fact, the previously known
Theorem~\ref{thm: hom-monoton} can also be shown along these lines.
We first outline such a proof and then describe the modification that
yields the new Theorem~\ref{thm: poly-mon}. The following filters
will be relevant:
\begin{enumerate}
\item A \emph{cardinality filter} allows us to restrict the hom-expansion
of $p$ to graphs with a desired number of vertices. Such a filter
can be constructed via polynomial interpolation arguments that are
standard in counting complexity.
\item Conditioned on the cardinality restriction, a \emph{color-surjectivity
filter} restricts the hom-expansion further to graphs that include
$S$ as a subgraph. Such filters can be realized using the inclusion-exclusion
principle, incurring a running time overhead of $2^{|E(S)|}$. Our
main contribution in this paper is an efficient and possibly surprising
construction based on \emph{Cai-Fürer-Immerman graphs}, which were
originally used in the context of graph isomorphism and finite model
theory~\cite{DBLP:journals/combinatorica/CaiFI92}.
\end{enumerate}
In the remainder of this section, we first introduce the general notion
of filters more formally. Then we elaborate on the two concrete filters
described above and sketch previously known constructions thereof.
Finally, we describe our new contribution of color-surjectivity filters
from CFI graphs. This section already covers large parts of the actual
proof in Section~\ref{sec: Filtering} and therefore introduces some
of the concepts and notation used in later sections.

\subsubsection*{General facts on filters}

Filters are implemented as \emph{quantum graphs}, which were introduced
by Lovász~\cite{Lovasz2012} in the context of graph algebras. They
can be treated like graphs, both in conceptual and computational terms.
More precisely, a\emph{ quantum graph }is a formal linear combination
\[
\mathbf{G}=\sum_{i=1}^{t}\alpha_{i}\,G_{i}
\]
with \emph{coefficients} $\alpha_{1},\ldots,\alpha_{t}\in\mathbb{Q}$
and \emph{constituent graphs} $G_{1},\ldots,G_{t}$, either all uncolored
or all colored. We assume the constituent graphs to be pairwise non-isomorphic,
as we could otherwise collect isomorphic graphs and represent them
by one single graph whose coefficient is the sum of the collected
coefficients. Graph parameters $p$ are linearly extended to quantum
graphs via
\[
p(\mathbf{G}):=\sum_{i=1}^{t}\alpha_{i}\,p(G_{i}).
\]

Based on these definitions, filters are quantum graphs $\mathbf{F}$
with the property that homomorphism counts into $\mathbf{F}$ recognize
the ``good'' graphs $\mathcal{G}$ among some set of ``filterable''
graphs $\mathcal{F}$. 
\begin{defn}
A quantum graph $\mathbf{F}$ \emph{filters} a set of graphs $\mathcal{G}$
out of a set of graphs $\mathcal{F}\supseteq\mathcal{G}$ if, for
all $H\in\mathcal{F}$,
\[
\hom(H,\mathbf{F})=\begin{cases}
1 & H\in\mathcal{G},\\
0 & H\in\mathcal{F}\setminus\mathcal{G}.
\end{cases}
\]
Note that no statement is made about graphs $H\notin\mathcal{F}$.
\end{defn}

\begin{example}
The quantum graph $\mathbf{F}=\frac{1}{24}K_{4}-\frac{1}{6}K_{3}$
filters $\mathcal{G}=\{K_{4}\}$ out of $\mathcal{F}=\{K_{3},K_{4}\}$,
since 
\begin{align*}
\hom(K_{4},\mathbf{F}) & =\frac{1}{24}24-\frac{1}{6}0=1,\\
\hom(K_{3},\mathbf{F}) & =\frac{1}{24}24-\frac{1}{6}6=0.
\end{align*}
No guarantees are made for graphs $H\notin\mathcal{F}$. For example,
we have $K_{2}\notin\mathcal{F}$ and 
\[
\hom(K_{2},\mathbf{F})=\frac{1}{24}12-\frac{1}{6}6=-\frac{1}{2}.
\]
\end{example}

Given a graph motif parameter $p$ on support $\mathcal{F}$ and a
quantum graph $\mathbf{F}$ that filters some set $\mathcal{G}$ out
of $\mathcal{F}$, graph products with $\mathbf{F}$ will allow us
to restrict the hom-expansion of $p$ to $\mathcal{G}$. These products
are defined as follows, see also~\cite[Chapter~3.3]{Lovasz2012}
for a variant involving \emph{uncolored graphs} instead of (possibly
colored) quantum graphs.
\begin{defn}
Given a graph $G$ with coloring $c:V(G)\to C$ and $i\in C$, write
$V_{i}(G)$ for the $i$-colored vertices in $G$. For graphs $G$
and $X$ with colors $C$, the \emph{tensor product} $G\otimes X$
is the graph on colors $C$ with 
\begin{itemize}
\item vertex set $V_{i}(G\otimes X)=V_{i}(G)\times V_{i}(X)$ for every
color $i\in C$, and 
\item an edge between $(u_{G},u_{X})$ and $(v_{G},v_{X})$ iff $u_{G}v_{G}\in E(G)$
and $u_{X}v_{X}\in E(X)$. 
\end{itemize}
For quantum graphs $\mathbf{G}=\sum_{i}\alpha_{i}G_{i}$ and $\mathbf{X}=\sum_{j}\beta_{j}X_{j}$,
we set $\mathbf{G}\otimes\mathbf{X}=\sum_{i,j}(\alpha_{i}\beta_{j})\,(G_{i}\otimes X_{j}).$
\end{defn}

A very useful algebraic relation holds for homomorphism counts and
tensor products, see~\cite[(5.30)]{Lovasz2012} for uncolored graphs.
\begin{fact}
\label{fact: tensor-mult}Given a graph $F$ and quantum graphs $\mathbf{G}$
and $\mathbf{X}$, either all colored or all uncolored, we have
\[
\hom(F,\mathbf{G}\otimes\mathbf{X})=\hom(F,\mathbf{G})\cdot\hom(F,\mathbf{X}).
\]
\end{fact}

\begin{proof}
We first prove the statement for graphs $G$ and $X$. Any homomorphism
$f:V(F)\to V(G\otimes X)$ induces a unique pair of homomorphisms
$(f_{G},f_{X})$ from $F$ to $G$ and $X$, where $f_{G}$ and $f_{X}$
are obtained by projecting all image vertices $f(v)$ for $v\in V(F)$
to the first or second component, respectively. Likewise, any such
pair $(f_{G},f_{X})$ induces a unique homomorphism from $F$ to $G\otimes X$
by pairing the image vertices $f_{G}(v)$ and $f_{X}(v)$ for all
$v\in V(F)$. The statement for quantum graphs then follows by distributivity.
\end{proof}
As a direct consequence of Fact~\ref{fact: tensor-mult}, we can
use tensor products with filters to restrict the hom-expansion of
a graph motif parameter from a set $\mathcal{F}$ to a desired set
$\mathcal{G}$. The following corollary encapsulates the result of
multiple applications of Fact~\ref{fact: tensor-mult}.
\begin{cor}
\label{cor:apply-filter}Let $p(\cdot)=\sum_{F\in\mathcal{F}}\alpha_{F}\hom(F,\pcdot)$
be a graph motif parameter for a finite set $\mathcal{F}$ and let
$\mathbf{F}_{1},\ldots,\mathbf{F}_{q}$ be quantum graphs with $k_{1},\ldots,k_{q}$
constituents of sizes $s_{1},\ldots,s_{q}$, that filter sets $\mathcal{G}_{1},\ldots,\mathcal{G}_{q}$
out of $\mathcal{F}$. Then $G\mapsto p(G\otimes\mathbf{F}_{1}\otimes\ldots\otimes\mathbf{F}_{q})$
satisfies $q(\cdot)=\sum_{F\in\mathcal{G}}\alpha_{F}\hom(F,\pcdot)$
for $\mathcal{G}=\bigcap_{i=1}^{q}\mathcal{G}_{i}$ and can be evaluated
on $n$-vertex graphs with $\prod_{i=1}^{q}k_{i}$ oracle calls to
$p$ on graphs with $n\prod_{i=1}^{q}s_{i}$ vertices.
\end{cor}

We remark that the idea of implementing desirable (but unobtainable)
graph functionalities by quantum graphs was used before by Lovász~\cite{Lovasz2012}
in the context of \emph{contractors} and \emph{connectors}, by Xia
and the author~\cite{Curticapean,Curticapean2015} when studying
the complexity of the permanent with \emph{combined matchgates}, and
by Dvor{\'{a}}k~\cite{DBLP:journals/jgt/Dvorak10} to characterize
the expressiveness of homomorphism counts.

\subsubsection*{Proving Theorem~\ref{thm: hom-monoton} via filters}

As a warm-up, let us prove the known Theorem~\ref{thm: hom-monoton}
by a combination of different filters. In the following, let $p$
be a graph motif parameter. Firstly, a cardinality filter allows us
to filter out graphs with exactly $k\in\mathbb{N}$ vertices from
its hom-expansion, subject to the condition that the filter is only
applied to graphs with at most $s\in\mathbb{N}$ vertices. We prove
the existence of more general filters in Section~\ref{sec: Filtering}.
\begin{fact}
\label{fact: first-card}Given integers $k,s\in\mathbb{N}$, there
is a quantum graph with $s+1$ constituents, each on $\leq s$ vertices,
that filters the graphs with exactly $k$ vertices out of the graphs
with $\leq s$ vertices. The constituents and coefficients (from $\mathbb{Q}$)
can be computed in polynomial time.
\end{fact}

Another straightforward construction allows us to filter $S$-colored
graphs out of all graphs: Given a colorful graph $S$, a graph $H$
is $S$\emph{-colored} if there is a homomorphism from $H$ to $S$.
Any colorful graph $S$ itself filters the $S$-colored graphs out
of all graphs. Less straightforwardly, we will also filter for graphs
that contain all edge-colors induced by the vertex-coloring of $S$: 
\begin{defn}
For a colorful graph $S$, an $S$-colored graph $H$ is \emph{surjectively
$S$-colored} if, for every edge $ij\in E(S)$, there is at least
one edge between vertices of colors $i$ and $j$ in $H$. We write
$E_{ij}(H)$ for the set of edges between vertices of colors $i$
and $j$ in $H$ and call $E_{ij}(H)$ an edge-color class of $H$.
\end{defn}

Besides $S$ itself, any graph obtained by repeatedly splitting off
edges from $S$ is surjectively $S$-colored: We may repeatedly remove
any edge $uv\in E_{ij}(S)$ for colors $i,j$ and add an edge between
fresh vertices $u^{*}$ and $v^{*}$ of colors $i$ and $j$ to obtain
an $S$-colored graph. This process can be repeated until left with
a vertex-colored matching on $2|E(S)|$ vertices. For surjectively
$S$-colored graphs $H$ with exactly $|V(S)|$ vertices, no such
splitting can be performed, and we obtain:
\begin{fact}
\label{fact: S-all-good}Let $S$ be a colorful graph without isolated
vertices. Any surjectively $S$-colored graph $H$ with $|V(S)|=|V(H)|$
satisfies $S\simeq H$.
\end{fact}

This fact is the reason why we wish to filter for graphs with a given
number of vertices, as in Fact~\ref{fact: first-card}. To implement
a filter for surjectively $S$-colored graphs, we can use the inclusion-exclusion
principle.
\begin{fact}
\label{fact: incl-excl}For every colorful graph $S$, the quantum
graph $\mathbf{I}=\sum_{F\subseteq S}(-1)^{|E(S)|-|E(F)|}F,$ with
$F$ ranging over the edge-subgraphs of $S$, filters the surjectively
$S$-colored graphs out of the set of all graphs.
\end{fact}

\begin{proof}
If $H$ is surjectively $S$-colored, then $\hom(H,\mathbf{I})=1$,
since $H$ has exactly one homomorphism to $S$ and none to proper
subgraphs. If $H$ is not surjectively $S$-colored, we show that
$\hom(H,\mathbf{I})=0$: Fix any edge $ij\in E(S)$ with $E_{ij}(H)=\emptyset$.
Any homomorphism from $H$ to a graph $F\subseteq S$ with $ij\notin E(F)$
is also a homomorphism from $H$ to $F+ij$, and vice versa. Since
$F$ and $F+ij$ have opposite signs in $\mathbf{I}$, their contributions
to $\hom(H,\mathbf{I})$ cancel out, so we obtain $\hom(H,\mathbf{I})=0$
overall.
\end{proof}
Finally, we combine the filters: Let $S$ be a colorful graph, let
$G$ be an input graph and let $p$ be a graph motif parameter that
contains $S$ and otherwise only graphs with $\leq s$ vertices. Using
the filters for cardinality (Fact~\ref{fact: first-card}) and surjectively
$S$-colored graphs (Fact~\ref{fact: incl-excl}), we obtain via
Corollary~\ref{cor:apply-filter} a graph motif parameter $q$ that
can be evaluated with $2^{|E(S)|}\mathrm{poly}(s)$ calls. By Fact~\ref{fact: S-all-good},
we have $q(\cdot)=\hom(S,\cdot)$, so we can compute $\hom(S,G)$
on $n$-vertex graphs $G$ in time $2^{|E(S)|}\mathrm{poly}(n,s)$
with an oracle for $p$. By a simple reduction that is described explicitly
in Section~\ref{sec: Filtering}, the above procedure can be performed
even when $p$ is defined on uncolored graphs and contains the uncolored
version $S^{\circ}$ instead of $S$ itself in its hom-expansion.

\subsubsection*{Efficient color-surjectivity filters from CFI graphs}

The main technical contribution of this paper is a new construction
of filters for surjectively colored graphs. The new filters are based
on graphs that were introduced by Cai, Fürer and Immerman~\cite{DBLP:journals/combinatorica/CaiFI92}
about 30 years ago to show limitations of the \emph{Weisfeiler-Lehman
method}~\cite{Weisfeiler1976OnCA,WeisfeilerLehman}, a particular
heuristic for graph isomorphism. The simplest instantiation of this
method is the so-called \emph{color refinement} algorithm, which compares
the iterated degree sequences of two graphs. This sequence is identical
for isomorphic graphs, but already very simple non-isomorphic graphs
may have identical iterated degree sequences. The \emph{$k$-dimensional}
Weisfeiler-Lehman ($k$-WL) algorithms for $k\in\mathbb{N}$ are higher-dimensional
variants of color refinement that compare iterated neighborhoods of
$k$-tuples of vertices. These algorithms have also been linked to
homomorphism counts~\cite{DBLP:conf/icalp/DellGR18,DBLP:journals/jgt/Dvorak10}
and were investigated in group-theoretic, algebraic and category-theoretic
settings~\cite{DBLP:conf/esa/BrachterS22,DBLP:conf/lics/AbramskyDW17,DBLP:journals/jsc/Derksen13}. 

\begin{figure}
\centering \def\svgwidth{12cm} 
\begingroup%
  \makeatletter%
  \providecommand\color[2][]{%
    \errmessage{(Inkscape) Color is used for the text in Inkscape, but the package 'color.sty' is not loaded}%
    \renewcommand\color[2][]{}%
  }%
  \providecommand\transparent[1]{%
    \errmessage{(Inkscape) Transparency is used (non-zero) for the text in Inkscape, but the package 'transparent.sty' is not loaded}%
    \renewcommand\transparent[1]{}%
  }%
  \providecommand\rotatebox[2]{#2}%
  \newcommand*\fsize{\dimexpr\f@size pt\relax}%
  \newcommand*\lineheight[1]{\fontsize{\fsize}{#1\fsize}\selectfont}%
  \ifx\svgwidth\undefined%
    \setlength{\unitlength}{234.60592195bp}%
    \ifx\svgscale\undefined%
      \relax%
    \else%
      \setlength{\unitlength}{\unitlength * \real{\svgscale}}%
    \fi%
  \else%
    \setlength{\unitlength}{\svgwidth}%
  \fi%
  \global\let\svgwidth\undefined%
  \global\let\svgscale\undefined%
  \makeatother%
  \begin{picture}(1,0.37084413)%
    \lineheight{1}%
    \setlength\tabcolsep{0pt}%
    \put(0,0){\includegraphics[width=\unitlength,page=1]{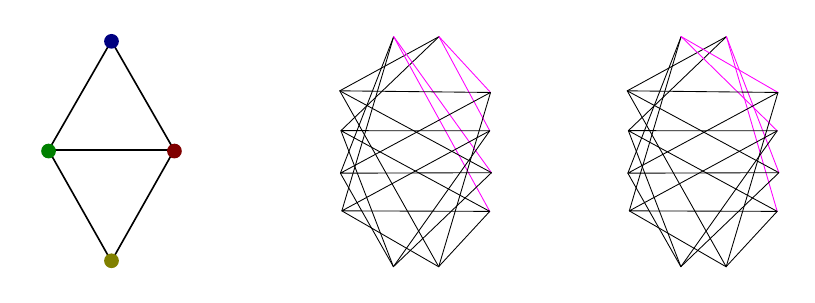}}%
    \put(0.06652885,0.33294184){\makebox(0,0)[lt]{\lineheight{1.25}\smash{\begin{tabular}[t]{l}$S$\end{tabular}}}}%
    \put(0.34175424,0.33591813){\makebox(0,0)[lt]{\lineheight{1.25}\smash{\begin{tabular}[t]{l}$X(S)$\end{tabular}}}}%
    \put(0.68972586,0.33591813){\makebox(0,0)[lt]{\lineheight{1.25}\smash{\begin{tabular}[t]{l}$\tilde X(S)$\end{tabular}}}}%
    \put(0,0){\includegraphics[width=\unitlength,page=2]{cfi.pdf}}%
  \end{picture}%
\endgroup%

\caption{\label{fig: CFI}A colorful graph $S$ and the CFI graphs $X(S)$
and $\tilde{X}(S)$. As indicated in the figure, for every $i\in V(S)$,
the $i$-colored vertices in $X(S)$ and $\tilde{X}(S)$ represent
even assignments in $\{0,1\}^{I(i)}$. For every edge $ij\in E(S)$,
an edge is present between two assignments in $X(S)$ if they agree
on their value for $ij$. The same is true for $\tilde{X}(S)$ except
that the pink edge-color class consisting of the edges between blue
and red vertices is complemented. The exceptional automorphism group
of $\tilde{X}(S)$ implies that the choice of edge-color class to
be complemented is \emph{irrelevant}: For all $ij\in E(S)$, complementing
$E_{ij}(X(S))$ yields the same graph $\tilde{X}(S)$ modulo isomorphism.}
\end{figure}

Cai, Fürer and Immerman~\cite{DBLP:journals/combinatorica/CaiFI92}
showed that, for every $k\in\mathbb{N}$, there exist two non-isomorphic
graphs that are not distinguished by $k$-WL. It follows that no fixed
level in the hierarchy of $k$-WL algorithms can solve the graph isomorphism
problem. By a known slight modification of the original CFI construction~\cite{roberson2022oddomorphisms,DBLP:journals/corr/abs-2304-07011},
we can transform any connected colorful graph $S$ into two $S$-colored
graphs $X=X(S)$ and $\tilde{X}=\tilde{X}(S)$, as shown in Figure~\ref{fig: CFI}.
 These graphs each have $2^{\Delta(S)-1}$ vertices per color class
and come with a remarkable property: For any edge $ij\in E(S)$, deleting
all edges between colors $i$ and $j$ renders $X$ and $\tilde{X}$
isomorphic. Intuitively speaking, this means that $X$ and $\tilde{X}$
can only be distinguished by inspecting all of their edge-color classes.
Thus, if some edge-color class of $S$ is missing in a graph $H$,
then homomorphisms from $H$ cannot distinguish $X$ and $\tilde{X}$,
since they cannot inspect this edge-color class, and we have $\hom(H,X)=\hom(H,\tilde{X})$
and thus $\hom(H,X-\tilde{X})=0$.\footnote{Similar features of CFI graphs were observed by Roberson~\cite{roberson2022oddomorphisms}
in the study of \emph{oddomorphisms} in an uncolored setting.} For the surjectively $S$-colored graph $S$ itself however, it can
be shown that $\hom(S,X-\tilde{X})=2^{|E(S)|-|V(S)|+1}$. We can thus
normalize by this number and obtain:
\begin{lem}
\label{lem: CFI filter}Given a colorful graph $S$ that is connected
and has no isolated vertices, there exists a quantum graph $\mathbf{X}$
with two constituents, each with at most $2^{\Delta(S)-1}$ vertices
per color class, that filters $\{S\}$ out of $\{S\}\cup\mathcal{N}(S)$,
where $\mathcal{N}(S)$ is the set of graphs that are not surjectively
$S$-colored. The constituents and coefficients of $\mathbf{X}$ can
be computed in time $O(4^{\Delta(S)}|V(S)|)$.
\end{lem}

Note that $\mathbf{X}$ is not actually a filter for surjectively
$S$-colored graphs, as it only filters $\{S\}$ out of $\{S\}\cup\mathcal{N}(S)$.
This will however suffice to prove Theorem~\ref{thm: poly-mon} along
the lines of Facts~\ref{fact: first-card} and \ref{fact: incl-excl}
if we replace the inclusion-exclusion based filter from Fact~\ref{fact: incl-excl}
by our new CFI-graph based filters from Lemma~\ref{lem: CFI filter}. 

Finally, in order to obtain the full Theorem~\ref{thm: poly-mon},
we need to reduce from $\hom(S,\pcdot)$ for colorful $S$ to graph
motif parameters $p$ with $T^{\circ}\inhom p$ for some supergraph
$T\supseteq S$. This requires adding ``dummy edges'', which yields
technicalities that lead to requirements (a)--(c) in the theorem
to avoid unwanted cancellations.

\subsection{Organization of the paper}

The general proof outline for Theorem~\ref{thm: poly-mon} was already
given in the introduction. The remainder of the paper contains five
sections: Section~\ref{sec:Preliminaries} collects preliminaries
on graph width measures and known hardness results for counting homomorphisms
from colorful graphs. In Section~\ref{subsec: color-surj filters},
we describe the CFI graph construction and prove Lemma~\ref{lem: CFI filter}.
The full proof of Theorem~\ref{thm: poly-mon} is given in Section~\ref{sec: Filtering},
and the theorem is used in Section~\ref{sec:Applications-of-Theorem}
to prove Theorems~\ref{thm: hard-hom}--\ref{thm: hard-ind}. In
Section~\ref{sec:Conclusion-and-future}, we conclude with an outlook
for future work based on this paper.

\subsection*{Acknowledgments}

The authors thanks Anuj Dawar, Pengming Wang, and David E.~Roberson
for conversations about CFI graphs that occurred over the past five
years.

\section{\label{sec:Preliminaries}Preliminaries}

In this paper, a \emph{counting problem} is a function $g:\{0,1\}^{*}\to\mathbb{Q}$.
The counting problem $\SAT$ asks, on input a binary encoding of a
Boolean formula $\varphi$, to count the satisfying assignments for
$\varphi$. We say that a counting problem $g$ is $\sharpP$\emph{-hard}
if there is a polynomial-time Turing reduction from $\SAT$ to $g$,
i.e., a polynomial-time algorithm that counts the satisfying assignments
of an input formula $\varphi$ when given oracle access to $g$. As
shown by Valiant~\cite{DBLP:journals/tcs/Valiant79}, the problem
of counting the perfect matchings in an input graph $G$ is $\sharpP$-hard.

\subsection*{Treewidth and walls}

Graphs $G$ will be undirected and may feature self-loops but no parallel
edges. As described in the introduction, they may be vertex-colored.
Given a vertex $v\in V(G)$, we write $I(v)$ for the set of edges
incident with $v$, assuming that $G$ is unique from the context.
The maximum degree of $G$ will be denoted by $\Delta(G)$. The treewidth
$\tw(G)$ of $G$ is a measure of ``tree-likeness'' that we do not
need to define formally in this paper. For a definition, see~\cite[Chapter 7]{Cygan2015}.

An \emph{elementary} $r\times r$ \emph{wall }for $r\in\mathbb{N}$
is the subgraph obtained from an $r\times r$ square grid by deleting
every odd-indexed edge from every odd-indexed column and every even-indexed
edge from every even-indexed column. An $r\times r$ \emph{wall} is
any graph obtained by subdividing some edges of an elementary $r\times r$
wall; note that its maximum degree is bounded by $3$. The \emph{wall
order} $\wall(G)$ of a graph $G$ is the maximum $r\in\mathbb{N}$
such that $G$ contains an $r\times r$ wall subgraph. By a highly
nontrivial result due to Chekuri and Chuzhoy~\cite{Chekuri2016},
the \emph{treewidth} $\tw(G)$ of $G$ is polynomially equivalent
to the largest number $s\in\mathbb{N}$ such that $G$ contains a
$s\times s$ square grid minor, and that number $s$ in turn is linearly
equivalent to the wall order.
\begin{thm}[\cite{Chekuri2016}]
\label{thm: find-wall}For any graph $G$ of treewidth $t\in\mathbb{N}$,
we have $\wall(G)\in\Omega(t^{1/100})$ and $\wall(G)\in O(t)$. Moreover,
there is a randomized polynomial-time algorithm that, given as input
a graph $G$ of treewidth $t$, outputs an $r\times r$ wall subgraph
of $G$ with $r\in\Omega(t^{1/100})$.
\end{thm}

\subsection*{Hardness of colorful homomorphisms\label{sec: hardness-colored}}

Given a class $\mathcal{H}$ of colorful graphs, let $\ColHom(\mathcal{H})$
be the problem of computing $\hom(H,G)$ on input graphs $H\in\mathcal{H}$
and $G$. Note that $\hom(H,G)$ refers to the number of color-preserving
homomorphisms from $H$ to $G$. For recursively enumerable graph
classes $\mathcal{H}$ of unbounded treewidth, the problem $\ColHom(\mathcal{H})$
is known to be $\sharpWone$-hard on parameter $|V(H)|$, as shown,
e.g., by the author and Marx~\cite{DBLP:conf/focs/CurticapeanM14},
who called the problem $\PartSub(\mathcal{H})$. That problem asks
to count color-preserving subgraph copies, but for colorful graphs
$H$, these correspond bijectively to color-preserving homomorphisms.

The $\sharpWone$-hardness of $\ColHom(\mathcal{H})$ was shown in
two steps, which can be adapted to obtain $\sharpP$-hardness of the
problem. First, let $\mathcal{G}$ denote the class of colorful square
grids. The proof of Theorem~3.2 in~\cite{DBLP:conf/focs/CurticapeanM14}
shows $\sharpWone$-hardness of $\ColHom(\mathcal{G})$ by a parameterized
reduction from counting $k$-cliques. Counting $k$-cliques is actually
$\sharpP$-hard, as the standard $\NP$-hardness reduction from Boolean
satisfiability preserves the number of solutions. The straightforward
reduction from counting $k$-cliques to $\ColHom(\mathcal{G})$ in
the proof of Theorem~3.2 in~\cite{DBLP:conf/focs/CurticapeanM14}
runs in polynomial time, so we obtain:
\begin{thm}
\label{thm:col-hom}The problem $\ColHom(\mathcal{G})$ for the class
$\mathcal{G}$ of colorful square grids is $\sharpP$-hard.
\end{thm}

By Theorem~\ref{thm: find-wall}, graphs of large treewidth contain
large square grids as minors. This can be used to reduce $\ColHom(\mathcal{G})$
to $\ColHom(\mathcal{H})$ for large-treewidth classes $\mathcal{H}$:
A \emph{minor model} of a graph $A$ in a graph $B$ is a family of
disjoint connected subsets $S_{v}\subseteq V(B)$ for each $v\in V(A)$
such that, for each edge $uv\in E(A)$, at least one edge runs between
$S_{u}$ and $S_{v}$. 
\begin{lem}[see proof of Lemma~3.1 in~\cite{DBLP:conf/focs/CurticapeanM14}]
\label{lem:col-minor}There is a polynomial-time algorithm to transform
an input graph $G$ and a minor model of a colorful graph $A$ in
a colorful graph $B$ into a graph $G'$ with $|V(G)|\cdot|V(B)|$
vertices such that $\hom(A,G)=\hom(B,G')$.
\end{lem}

We remark that proving hardness for the colorful version of counting
homomorphisms is somewhat easier than for the uncolored problem. As
discussed in the introduction, the known techniques to remove colors
require exponential running time in $H$.

\section{\label{subsec: color-surj filters}Constructing color-surjectivity
filters}

The constructions and arguments in this section are essentially due
to Cai, Fürer and Immerman~\cite{DBLP:journals/combinatorica/CaiFI92},
up to slight modifications that also already appeared in previous
works~\cite{roberson2022oddomorphisms,DBLP:journals/corr/abs-2304-07011}.
See Figure~\ref{fig: CFI} for an overview. The graphs constructed
in this section are encodings of constraint satisfaction problems
(CSPs) that are in turn derived from a graph $S$. For our purposes,
given a fixed colorful graph $S$, a \emph{CSP} is a pair $\Gamma=(\{A_{v}\}_{v\in V(S)},\{R_{uv}\}_{uv\in E(S)})$
consisting of
\begin{itemize}
\item a set of assignments $A_{v}\subseteq\{0,1\}^{I(v)}$ for each $v\in V(S)$,
and 
\item a relation $R_{uv}\subseteq A_{u}\times A_{v}$ for each $uv\in E(S)$.
\end{itemize}
A \emph{solution} of $\Gamma$ is a family of assignments $\{a_{v}\}_{v\in V(S)}$
such that $(a_{u},a_{v})\in R_{uv}$ for all $uv\in E(S)$. Slightly
abusing notation, we will also view $\Gamma$ as an $S$-colored graph
with $V_{v}(\Gamma)=A_{v}$ for $v\in V(S)$ and $E_{uv}(\Gamma)=R_{uv}$
for $uv\in E(S)$, and we ignore edge-directions in the relations.
The following is immediate from the definition of homomorphisms.
\begin{fact}
\label{fact: num-soln}The number of solutions to $\Gamma$ equals
$\hom(S,\Gamma)$.
\end{fact}

CFI graphs are constructed from particular CSPs in which $A_{v}$
for $v\in V(S)$ are the \emph{even} assignments in $\{0,1\}^{I(v)}$,
i.e., those assignments with an even numbers of $1$s. Given a ``charge
function'' $c:E(S)\to\{0,1\}$, we then define $\Gamma(S,c)$ by
setting, for all $uv\in E(S)$,
\begin{align*}
R_{uv}^{c} & =\{(a_{u},a_{v})\in A_{u}\times A_{v}\mid a_{u}(uv)\equiv_{2}a_{v}(uv)+c(uv)\},
\end{align*}
with equivalence modulo $2$. In other words, if $c(uv)=0$, then
$R_{uv}^{c}$ requires assignments $a_{u}$ and $a_{v}$ to agree
on their value for $uv$. If $c(uv)=1$, then $R_{uv}^{c}$ requires
them to disagree on this value. Given a set of edges $F\subseteq E(S)$,
let us introduce the notation $\chi_{F}\in\{0,1\}^{E(S)}$ to describe
the assignment that is $1$ on $F$ and $0$ otherwise. We also abbreviate
$\chi_{F}$ as $\chi_{e}$ and $\chi_{e,e'}$ when $|F|\leq2$. Then
we choose an arbitrary edge $e^{*}\in E(S)$ and define the quantum
graph 
\begin{equation}
\mathbf{X}(S)=\frac{\Gamma(S,\chi_{\emptyset})-\Gamma(S,\chi_{e^{*}})}{2^{|E(S)|-|V(S)|+1}}.\label{eq: def-of-XS}
\end{equation}
The CSP interpretation allows us to determine the number of homomorphisms
into $\mathbf{X}(S)$ explicitly.
\begin{lem}
\label{lem: homS-toXS}For any connected graph $S$, we have $\hom(S,\mathbf{X}(S))=1$.
\end{lem}

\begin{proof}
By Fact~\ref{fact: num-soln}, the value $\hom(S,\mathbf{X}(S))$
is the difference of the solution counts for the two constituents
$\Gamma(S,\chi_{\emptyset})$ and $\Gamma(S,\chi_{e^{*}})$, up to
normalization by $2^{|E(S)|-|V(S)|+1}$. We determine these solution
counts:
\begin{lyxlist}{00.00.0000}
\item [{$\Gamma(S,\chi_{\emptyset}):$}] The solutions for this CSP correspond
bijectively to \emph{even spanning subgraphs} of $S$, i.e., to edge-subsets
$B\subseteq E(S)$ such that each vertex $v\in V(S)$ has even degree
in $B$. As $S$ is connected, there are precisely $2^{|E(S)|-|V(S)|+1}$
such edge-subsets, as these form the \emph{cycle space} of $S$, which
is a subspace of $\mathbb{Z}_{2}^{E(S)}$ of dimension $|E(S)|-|V(S)|+1$.
\item [{$\Gamma(S,\chi_{e^{*}}):$}] We show that this CSP has no solutions.
Choose $u$ arbitrarily as one of the endpoints of $e^{*}$. The solutions
for $\Gamma(S,\chi_{e^{*}})$ correspond bijectively to edge-subsets
$B\subseteq E(S)$ such that each vertex $v\in V(S)$ has even degree
in $B$, except for $u$, which has odd degree. By the graph handshaking
lemma, no such $B$ exists, as it would induce a graph with precisely
one vertex of odd degree.
\end{lyxlist}
Combining the above, the lemma follows from the definition of $\mathbf{X}(S)$.
\end{proof}
Given functions $c$ and $c'$ with range $\{0,1\}$, let $c\oplus c'$
be their point-wise sum modulo $2$. We also write $c\oplus c'$ when
the domains $D$ and $D'$ of $c$ and $c'$ disagree and then define
the domain of $c\oplus c'$ as $D\cap D'$. Two CSPs are declared
to be isomorphic if they are isomorphic as colored graphs. The following
lemma shows an extremely useful property of CFI graphs: Given a charge
function $c\in\{0,1\}^{E(S)}$ and two incident edges $vu,vw\in E(S)$,
we can flip $c$ at $vu$ and $vw$ and obtain a new charge $c'$
that yields an isomorphic CFI graph.
\begin{lem}
\label{lem: push-incident}For all $c\in\{0,1\}^{E(S)}$ and $vu,vw\in E(S)$,
we have $\Gamma(S,c)\simeq\Gamma(S,c\oplus\chi_{vu,vw})$.
\end{lem}

\begin{proof}
We define an isomorphism $f$ from $\Gamma(S,c)$ to $\Gamma(S,c\oplus\chi_{vu,vw})$
by setting
\[
f(a)=\begin{cases}
a\oplus\chi_{vu,vw} & a\in A_{v},\\
a & a\in A_{z}\text{ with }z\neq v.
\end{cases}
\]
As any $a\in A_{z}$ for $z\in V(S)$ is even and its image $f(a)$
is obtained by flipping two (if $z=v)$ or zero (if $z\neq v$) entries,
this image is also even and thus contained in $A_{z}$. The function
$f$ is therefore well-defined.

To check that $f$ preserves the edge-relation, let us abbreviate
$c'=c\oplus\chi_{vu,vw}$ and $a'=f(a)$ for all $a\in V(\Gamma)$.
Then, for $z\in\{v,w\}$, we have $(a_{u},a_{z})\in R_{uz}^{c}$ iff
$(a_{u}',a_{z}')\in R_{uz}^{c'}$. This is because $a_{u}(uz)=a_{z}(uz)$
iff $a'_{u}(uz)\neq a'_{z}(uz)$, while all other relations are trivially
preserved. This shows that $f$ is indeed an isomorphism from $\Gamma(S,c)$
to $\Gamma(S,c\oplus\chi_{vu,vw})$.
\end{proof}
Using this, we can ``push'' charge from any edge to another edge
in the same connected component.
\begin{lem}
\label{lem: edge-push}Let $e,e'\in E(S)$. If $S$ is connected,
then $\Gamma(S,\chi_{e})\simeq\Gamma(S,\chi_{e'})$.
\end{lem}

\begin{proof}
Let $P=(e,\ldots,e')$ be a sequence of edges in $S$ in which consecutive
edges share a common endpoint; this sequence exists because $S$ is
connected. The required isomorphism follows from repeated applications
of Lemma~\ref{lem: push-incident} along consecutive edges in $P$.
\end{proof}
We can use this lemma to show that ``every edge-color counts'' in
the CFI construction, in the sense that the constituents of $\mathbf{X}(S)$
become isomorphic when the edges between any two vertex-color classes
are deleted. Given an $S$-colored graph $F$ and a proper subgraph
$S'\subseteq S$ of the colorful graph $S$, note that the tensor
product $F\otimes S'$ is canonically isomorphic to the graph obtained
from $F$ by deleting the edge-color classes corresponding to edges
in $E(S)\setminus E(S')$. Mildly abusing notation, we view $F\otimes S'$
as an actual subgraph of $F$.
\begin{cor}
\label{cor: del-to-iso}For any proper subgraph $S'$ of $S$, we
have $\Gamma(S,\chi_{e^{*}})\otimes S'\simeq\Gamma(S,\chi_{\emptyset})\otimes S'$.
\end{cor}

\begin{proof}
Let $e\in E(S)\setminus E(S')$ with $e=uv$. By Lemma~\ref{lem: edge-push},
we can ``push'' the charge from $e^{*}$ to $e$, i.e., we have
$\Gamma(S,\chi_{e^{*}})\simeq\Gamma(S,\chi_{e})$. This also implies
$\Gamma(S,\chi_{e^{*}})\otimes S'\simeq\Gamma(S,\chi_{e})\otimes S'$,
because deleting entire vertex-color and edge-color classes from two
isomorphic colored graphs preserves their isomorphism. On the other
hand, we have $\Gamma(S,\chi_{e})\otimes S'\simeq\Gamma(S,\chi_{\emptyset})\otimes S'$
by the trivial identity isomorphism, because $\Gamma(S,\chi_{e})$
and $\Gamma(S,\chi_{\emptyset})$ differ only on edges in $R_{uv}$.
Composing these two isomorphisms yields the claim.
\end{proof}
Finally, we collect the facts shown in this section to prove Lemma~\ref{lem: CFI filter}.
\begin{proof}[Proof of Lemma~\ref{lem: CFI filter}]
 Given a connected graph $S$, define the quantum graph $\mathbf{X}(S)$
as in (\ref{eq: def-of-XS}). Viewed as $S$-colored graphs, the CSPs
$\Gamma(S,\chi_{\emptyset})$ and $\Gamma(S,\chi_{e^{*}})$ satisfy
the requirements on the color class sizes and can be computed in the
required time. Note that up to $O(4^{\Delta(S)})$ time may be required
to write down all edges in a given edge-color class of the two graphs.

Lemma~\ref{lem: homS-toXS} shows that $\hom(S,\mathbf{X})=1$. To
prove that $\hom(H,\mathbf{X})=0$ for graphs $H$ that are not surjectively
$S$-colored, let $H$ be such a graph. We may assume that $H$ is
$S$-colored, as otherwise $\hom(H,\mathbf{X})=0$ because the constituents
of $\mathbf{X}$ are $S$-colored. It follows that $H$ is $S'$-colored
for $S'\subsetneq S$, and we obtain 
\[
\hom(T,\mathbf{X}(S))=\hom(T,\mathbf{X}(S)\otimes S')=0,
\]
where the first equality holds because the image of $T$ in any constituent
$X$ of $\mathbf{X}(S)$ is contained in $X\otimes S'$, and the second
equality holds because the two constituents of $\mathbf{X}(S)\otimes S'$
are isomorphic by Corollary~\ref{cor: del-to-iso}. This proves the
lemma.
\end{proof}

\section{\label{sec: Filtering}Filtering a linear combination}

In this section, we prove Theorem~\ref{thm: poly-mon} along the
outline described in Section~\ref{subsec: proof-ideas} of the introduction:
Given a set $\mathcal{H}$ of pairwise non-isomorphic uncolored graphs
and access to a graph parameter 
\[
p(\cdot)=\sum_{H\in\mathcal{H}}\alpha_{H}\hom(H,\pcdot),
\]
we wish to determine $\hom(S,G)$ for a colorful graph $S$ and a
colored graph $G$, both on color set $C$.

First, we need to address the conversion from colored to uncolored
graphs: Recall that we write $G^{\circ}$ for the uncolored graph
obtained from $G$ by ignoring colors. We write $\mathbf{G}^{\circ}$
for the quantum graph obtained by applying this operator constituent-wise.
Moreover, we write $H^{c}$ for the colored graph obtained from $H$
by applying $c$ as coloring.
\begin{lem}
\label{lem: col2uncol}Let $C$ be a set of colors and let $p_{\mathrm{col}}$
denote the graph motif parameter on colored graphs obtained from the
uncolored graph parameter $p$ via 
\begin{equation}
p_{\mathrm{col}}(\cdot)=\sum_{H\in\mathcal{H}}\alpha_{H}\sum_{c:V(H)\to C}\hom(H^{c},\pcdot).\label{eq: p-color-subst}
\end{equation}
Given a colored graph $G$, we have $p(G^{\circ})=p_{\mathrm{col}}(G).$
After collecting for isomorphic terms, the coefficient of any colorful
graph $F$ in $p_{\mathrm{col}}$ equals $\alpha_{F^{\circ}}|\aut(F^{\circ})|$.
\end{lem}

Note that the right-hand side of (\ref{eq: p-color-subst}) may contain
colorful graphs that are not colored by the identity function.
\begin{proof}
We first show $p(G^{\circ})=p_{\mathrm{col}}(G)$ when $p(\cdot)=\hom(H,\pcdot)$
for an uncolored graph $H$. The general claim follows by linearity.
Given a graph $G$ on colors $C$, we have $\hom(H,G^{\circ})=\sum_{c:V(H)\to C}\hom(H^{c},G)$,
because any homomorphism $f$ from $H$ into $G^{\circ}$ induces
a unique coloring $c_{f}$ of $H$ via its image: For $v\in V(H)$,
let $c_{f}(H)$ be the color of $f(v)\in V(G)$. The set of homomorphisms
from $H$ to $G^{\circ}$ can be partitioned according to these colorings,
and the homomorphisms inducing a given coloring $c:V(H)\to C$ are
precisely the homomorphisms from $H^{c}$ to $G$.

For the second statement, let $F$ be a colorful graph. Every pair
$(F^{\circ},c)$ with $F\simeq(F^{\circ})^{c}$ induces an automorphism
of $F^{\circ}$ via the bijection $c:V(F^{\circ})\to C$.
\end{proof}
Our proof relies on a general version of the cardinality filter discussed
in the introduction.
\begin{defn}
Consider a partition of a color set $C$ into $r\in\mathbb{N}$ parts
$C_{1},\ldots,C_{r}$ that are annotated with numbers $k_{1},\dots,k_{r}\in\mathbb{N}$.
We call $\eta=(C_{1},\ldots,C_{r},k_{1},\ldots,k_{r})$ a \emph{color-coarsening}
of $C$ with $r$ parts and say that $H$ is $\eta$\emph{-coarsened}
if $\sum_{j\in C_{i}}|V_{j}(H)|=k_{i}$ holds for all $i\in[r]$. 
\end{defn}

That is, if $H$ is $\eta$-coarsened, then for every $i\in[r]$,
there are exactly $k_{i}$ vertices in $H$ with colors from $C_{i}$.
In particular, a graph $H$ on color set $C$ has $k\in\mathbb{N}$
vertices iff it is $(C,k)$-coarsened.
\begin{lem}
\label{lem: number-filter}Let $\eta$ be a color-coarsening with
$r\in\mathbb{N}$ parts of a color set $C$. For all $s\in\mathbb{N}$,
there exists a quantum graph $\mathbf{N}$ that filters the $\eta$-coarsened
graphs out of the graphs with $\leq s$ vertices. The coefficients
and constituents of $\mathbf{N}$ can be computed in $s^{O(r)}$ time
on input $\eta$.
\end{lem}

\begin{proof}
Let $\eta=(C_{1},\ldots,C_{r},k_{1},\ldots,k_{r})$. Given $a=(a_{1},\ldots,a_{r})\in[s+1]^{r}$,
define a graph $N_{a}$: For $i\in[r]$, this graph contains exactly
$a_{i}$ vertices of color $j$, for all colors $j\in C_{i}$. All
vertices in $N_{a}$ have self-loops and edges to all other vertices.
Given a colored graph $H$ with colors $C$, write $n_{i}(H)=\sum_{j\in C_{i}}|V_{j}(H)|$
for $i\in[r]$. For all $a\in[s+1]^{r}$, we have
\begin{equation}
\hom(H,N_{a})=\prod_{i\in[r]}a_{i}^{n_{i}(H)}.\label{eq: homGNAB-1}
\end{equation}
For any fixed graph $H$, the right-hand side can be viewed as a multivariate
polynomial $p_{H}\in\mathbb{Q}[a_{1},\ldots,a_{r}]$ with maximum
degree $s$. This polynomial has exactly one monomial, and this monomial
is $a_{1}^{k_{1}}\ldots a_{r}^{k_{r}}$ iff $H$ is $\eta$-coarsened.
Thus, the coefficient $c_{k_{1},\ldots,k_{r}}$ of $a_{1}^{k_{1}}\ldots a_{r}^{k_{r}}$
in $p_{H}$ is $1$ if $H$ is $\eta$-coarsened, and it is $0$ otherwise.
By multivariate polynomial interpolation, there are coefficients $\beta_{a}$
for $a\in[s+1]^{r}$ such that
\[
c_{k_{1},\ldots,k_{r}}=\sum_{a\in[s+1]^{r}}\beta_{a}p_{H}(a)=\sum_{a\in[s+1]^{r}}\beta_{a}\hom(H,N_{a}),
\]
where the coefficients $\beta_{a}$ can be computed in polynomial
time by relating the evaluations and coefficients of $p_{H}$ through
a full-rank linear system of equations on $(s+1)^{r}$ indeterminates.
It follows that $\mathbf{N}:=\sum_{a\in[s+1]^{r}}\beta_{a}N_{a}$
satisfies the requirements of the lemma.
\end{proof}
To prove Theorem~\ref{thm: poly-mon}, we invoke Lemma~\ref{lem: CFI filter}
from Section~\ref{subsec: color-surj filters} together with Lemma~\ref{lem: number-filter}
on $r\leq2$ and $s=\max_{H\inhom p}|V(H)|$. It will not suffice
to choose $r=1$, since we may need to distinguish between ``interesting''
and ``dummy'' vertices in parts of the proof.
\begin{proof}[Proof of Theorem~\ref{thm: poly-mon}]
We may assume that $S$ is connected: If $S$ has connected components
$S_{1},\ldots,S_{q}$ with color sets $C_{1},\ldots,C_{q}$ for $q\geq2$,
then we have $\hom(S,G)=\prod_{i=1}^{q}\hom(S_{i},G_{i})$, where
$G_{i}$ is the restriction of $G$ to vertices of color $C_{i}$,
so it suffices to compute $\hom(S_{i},G_{i})$ for $i\in[q]$. Assuming
then that $S$ is connected, let $\mathbf{X}=\mathbf{X}(S)$ be the
quantum graph from Lemma~\ref{lem: CFI filter}. Its two constituents
on $\leq2^{\Delta(S)}|V(S)|$ vertices and coefficients can be computed
in $O(4^{\Delta(S)}|V(S)|)$ time.

We first describe the reduction in Case~(a). Lemma~\ref{lem: number-filter}
yields a quantum graph $\mathbf{N}$ that filters the $(C,|V(S)|)$-coarsened
graphs out of the graphs with $\leq s$ vertices. The $\mathrm{poly}(s)$
constituents and coefficients of $\mathbf{N}$ can be computed in
$\mathrm{poly}(s)$ time. The $(C,|V(S)|)$-coarsened graphs are precisely
the graphs on color set $C$ with $|V(S)|$ vertices. By Fact~\ref{fact: S-all-good},
the surjectively $S$-colored graphs among those graphs are isomorphic
to $S$. Using Corollary~\ref{cor:apply-filter} with $\mathbf{X}$
and $\mathbf{N}$ on $p_{\mathrm{col}}$, we can thus evaluate $\alpha_{S^{\circ}}|\aut(S^{\circ})|\cdot\hom(S,G)$
with $2^{\Delta(S)}\mathrm{poly}(s)$ oracle calls of the form $p_{\mathrm{col}}(R)=p(R^{\circ})$.
Since $\hom(S,S)=1$, we can moreover compute $\alpha_{S^{\circ}}|\aut(S^{\circ})|\neq0$
by invoking the above steps with $G=S$. This allows us to compute
$\hom(S,G)$.

In Case~(b), we proceed similarly, but we add ``padding'' to the
host graphs to capture the surplus edges in $T\supseteq S$. That
is, for any $S$-colored graph $G$, we define a graph $G_{\mathrm{pad}}$
as follows, starting from $G$:
\begin{enumerate}
\item For every vertex $v\in V(T)\setminus V(S)$, add a vertex with the
same color as $v$ to $G_{\mathrm{pad}}$.
\item For every edge $ij\in E(T)\setminus E(S)$, add all edges $uv$ between
colors $i$ and $j$ in $G_{\mathrm{pad}}$.
\end{enumerate}
Note that $G_{\mathrm{pad}}$ is $T$-colored. For any $T$-colored
graph $H$, consider its restriction $H\otimes S$ to vertex- and
edge-colors from $S$. Any homomorphism from $H$ to $G_{\mathrm{pad}}$
can be restricted to a homomorphism from $H\otimes S$ to $G$ by
deleting the image vertices and edges not contained in $G$. Moreover
there exists a number $c_{H}\neq0$, independent of $G$, such that
any homomorphism $f$ from $H\otimes S$ to $G$ admits exactly $c_{H}$
homomorphisms from $H$ to $G_{\mathrm{pad}}$ that restrict to $f$.
That is,
\begin{equation}
\hom(H,G_{\mathrm{pad}})=c_{H}\cdot\hom(H\otimes S,G).\label{eq: Gpad}
\end{equation}
Let $C_{S}\subseteq C_{T}$ denote the colors of $S$ and $T$. We
set $\eta=(C_{S},C_{T}\setminus C_{S},|V(S)|,|V(T)|-|V(S)|)$ and
write $\mathcal{C}$ for the set of $T$-colored and $\eta$-coarsened
graphs $H^{c}$ with $H\in\mathcal{H}$ and $c:V(H)\to C$ such that
$H^{c}\otimes S$ is surjectively $S$-colored. Note that $\mathcal{C}$
may contain pairs of isomorphic graphs. In fact, since $|V(H\otimes S)|=|V(S)|$
for all $H\in\mathcal{C}$, it follows by Fact~\ref{fact: S-all-good}
that 
\begin{equation}
H\otimes S\simeq S\quad\text{for all }H\in\mathcal{C}.\label{eq: Srestrict}
\end{equation}
Using Corollary~\ref{cor:apply-filter} with $\mathbf{X}$ and $\mathbf{N}$
on $p^{\mathrm{col}}$, we can thus use $2^{\Delta(S)}\mathrm{poly}(s)$
oracle calls to evaluate
\[
\sum_{H\in\mathcal{C}}\alpha_{H^{\circ}}\hom(H,G_{\mathrm{pad}})\:=\ \sum_{H\in\mathcal{C}}c_{H}\alpha_{H^{\circ}}\hom(H\otimes S,G)\:=\ \sum_{H\in\mathcal{C}}c_{H}\alpha_{H^{\circ}}\hom(S,G),
\]
where the equalities are due to (\ref{eq: Gpad}) and (\ref{eq: Srestrict}).
Note that the last sum ranges over $H\in\mathcal{C}$ but only involves
homomorphism counts from $S$. Since all graphs $H\in\mathcal{C}$
have exactly $|V(T)|$ vertices, their coefficients $\alpha_{H^{\circ}}$
all have the same signs by the assumptions of Case~(b), so it follows
that $q_{\mathcal{C}}:=\sum_{H\in\mathcal{C}}c_{H}\alpha_{H^{\circ}}\neq0$.
As in Case~(a), we can then determine $q_{\mathcal{C}}$ with additional
oracle calls by setting $G=S$ and finally compute $\hom(S,G)$.

Case~(c) is shown similarly to Case~(b): We construct $G_{\mathrm{pad}}$,
define $\mathcal{C}$, and evaluate $\sum_{H\in\mathcal{C}}c_{H}\alpha_{H^{\circ}}\hom(S,G)$
as above. All graphs in $\mathcal{C}$ are $T$-colored and have exactly
$|V(T)|$ vertices, so they are edge-subgraphs of $T$. By the assumptions
of Case~(c), this only leaves recolored versions of $T$ in $\mathcal{C}$,
and thus $q_{\mathcal{C}}=\sum_{H\in\mathcal{C}}c_{H}\alpha_{H^{\circ}}\neq0$
and we can proceed as in Case~(b) to compute $\hom(S,G)$.
\end{proof}

\section{\label{sec:Applications-of-Theorem}Applications of the main reduction}

In this section, we show Theorems~\ref{thm: hard-hom} and \ref{thm: hard-sub}
as applications of Theorem~\ref{thm: poly-mon}. We start with the
result for $\Hom(\mathcal{H})$, the problem of counting uncolored
homomorphism counts from a fixed class of patterns $\mathcal{H}$.
Our new theorem translates a fundamental $\sharpWone$-hardness result
in parameterized complexity by Dalmau and Jonsson~\cite{Dalmau2004}
to a $\sharpP$-hardness result.
\begin{proof}[Proof of Theorem~\ref{thm: hard-hom}]
We show $\sharpP$-hardness of $\Hom(\mathcal{H})$ by reduction
from $\ColHom(\mathcal{G})$ for the class of colorful grids, which
is $\sharpP$-hard by Theorem~\ref{thm:col-hom}. As input for $\ColHom(\mathcal{G})$,
let $H\in\mathcal{G}$ be a colorful grid and let $G$ be a colored
graph. Using the assumption of the theorem, we can compute a colorful
graph $T$ with $T^{\circ}\in\mathcal{H}$ in time $\mathrm{poly}(|V(H)|)$
such that $S\subseteq T$ for a wall $S$ that is large enough to
contain $H$ as a minor. We either use Theorem~\ref{thm: find-wall}
to find $S$ in randomized polynomial time or use the deterministic
algorithm to find it.

Using Lemma~\ref{lem:col-minor}, we then compute a graph $G'$ with
$\hom(H,G)=\hom(S,G')$ in polynomial time. Using Theorem~\ref{thm: poly-mon},
we then determine $\hom(S,G')$ in polynomial time with an oracle
for $\hom(T^{\circ},\pcdot)$: Indeed, we have $\Delta(S)\leq3$ and
$T^{\circ}$ is the only graph appearing in the hom-expansion of $\hom(T^{\circ},\pcdot)$,
so the sign condition in Case~(b) in Theorem~\ref{thm: poly-mon}
is trivially fulfilled, yielding the desired reduction to $\Hom(\mathcal{H})$.
\end{proof}
We continue with the subgraph counting problems $\Sub(\mathcal{H})$.
Significant technical effort was required to obtain $\sharpWone$-hardness
for $\Sub(\mathcal{H})$ in the original paper by Marx and the author~\cite{DBLP:conf/focs/CurticapeanM14},
and even the machinery developed in follow-up works~\cite{DBLP:conf/stoc/CurticapeanDM17}
did not suggest that a general $\sharpP$-hardness result was in reach.
\begin{proof}[Proof of Theorem~\ref{thm: hard-sub}]
By Theorem~\ref{thm:col-hom} and Lemma~\ref{lem:col-minor}, the
problem $\ColHom(\mathcal{W})$ for the class of colorful walls is
$\sharpP$-hard. We reduce it to $\Sub(\mathcal{H})$. Let $S\in\mathcal{W}$
and a graph $G$ be given as input for $\ColHom(\mathcal{W})$. Using
the assumption of the theorem, we find a graph $H\in\mathcal{H}$
and a matching $M$ in $H$ with $|E(S)|$ edges in polynomial time.
By identifying the endpoints of edges in $M$, any graph on $|E(S)|$
edges can be found in polynomial time as a subgraph of some quotient
of $H$. In particular, there is a colorful graph $T\supseteq S$
such that $T^{\circ}$ is a quotient of $H$. This graph $T$ and
the subgraph copy of $S$ in $T$ can be computed in polynomial time.

We use Theorem~\ref{thm: poly-mon} to compute $\hom(S,G)$ in polynomial
time with an oracle for $\sub(H,\pcdot)$: Note that $\Delta(S)\leq3$
and that all graphs $F\inhom\sub(H,\pcdot)$ have at most $|V(H)|$
vertices. By Fact~\ref{fact: sub hom-exp}, we have $T^{\circ}\inhom\sub(H,\pcdot)$,
since $T^{\circ}$ is a quotient of $H$, and all graphs $R$ with
$|V(R)|=|V(T)|$ have the same sign in the hom-expansion. Invoking
Theorem~\ref{thm: poly-mon} with Case~(b) yields the desired reduction
from $\ColHom(\mathcal{W})$ to $\Sub(\mathcal{H})$.
\end{proof}
We conclude with a proof for the induced subgraph counting problems
$\Ind(\mathcal{H})$.
\begin{proof}[Proof of Theorem~\ref{thm: hard-ind}]
For all $t\in\mathbb{N}$, we can compute a graph $H_{t}$ with $|V(H_{t})|\geq t$
in polynomial time. At least one of $H_{t}$ or its complement has
average degree $\Omega(t)$, and we may assume that $H_{t}$ has,
as we could otherwise complement $H_{t}$ and the host graphs without
changing the induced subgraph count. Using known algorithms~\cite[Fact 7]{DBLP:journals/tcs/AlonLW07},
we can then find a minor model of $K_{q}$ in $H_{t}$ for $q\in\Omega(t/\sqrt{\log t})$
in polynomial time. From this minor model, we can then find a $q'\times q'$
wall subgraph $S$ with $q'\in\Omega(t^{1/2-\epsilon})$ in $H_{t}$.

By Theorem~\ref{thm:col-hom} and Lemma~\ref{lem:col-minor}, the
problem $\ColHom(\mathcal{E})$ for the class $\mathcal{E}$ of colorful
elementary walls is $\sharpP$-hard. Let $W\in\mathcal{E}$ and let
$G$ be a graph. As in the previous paragraph, we can find a graph
$H_{t}\in\mathcal{H}$ such that $S^{\circ}\subseteq H_{t}$ for a
subdivision $S$ of $W$. Lemma~\ref{lem:col-minor} allows us to
construct a graph $G'$ such that $\hom(W,G)=\hom(S,G')$. By Fact~\ref{fact: ind hom-exp},
the graphs $F\inhom\ind(H_{t},\pcdot)$ have at most $|V(H_{t})|$
vertices, and among those $F$ in the hom-expansion with $|V(F)|=|V(H_{t})|$,
the graph $H_{t}$ itself is edge-minimal. We can therefore invoke
Theorem~\ref{thm: poly-mon} in Case~(c) to compute $\hom(W',G')$
in polynomial time with an oracle for $\ind(H_{t},\pcdot)$.
\end{proof}

\section{\label{sec:Conclusion-and-future}Conclusion and future work}

We proved $\sharpP$-hardness for certain families of graph parameters
$(p_{t})_{t\in\mathbb{N}}$ in which each parameter $p$ is a linear
combination
\[
p(\cdot)=\sum_{F}\alpha_{F}\hom(F,\pcdot)
\]
with finitely many terms, and the expansions contain polynomial-time
constructible graphs of polynomial treewidth. First, we proved $\sharpP$-hardness
of counting colorful homomorphisms from the high-treewidth graphs
appearing in the hom-expansions. Then we used tensor products with
quantum graphs to narrow down the hom-expansion of $p$ to $\hom(S,\pcdot)$
for some high-treewidth graph $S$. Combined with known combinatorial
results that characterize graphs $S\inhom p$, this reduction yields
the desired $\sharpP$-hardness results. 

There is potential for interesting follow-up works: Most importantly,
it is likely that the requirements in Theorem~\ref{thm: poly-mon}
can be weakened. How close can we get to merely requiring $T^{\circ}\inhom p$
for some supergraph $T\supseteq S$, without imposing additional requirements
on the coefficients of other graphs besides $T^{\circ}$? It may not
even be necessary to provide $T$ as an input, as it may be possible
to find $T$ using the oracle for $p$.

Following up on the initial paper on the complexity monotonicity framework~\cite{DBLP:conf/stoc/CurticapeanDM17},
involved techniques from algebra and group theory were used to identify
large-treewidth graphs $S$ with $S\inhom p$ for interesting graph
motif parameters $p$~\cite{DBLP:conf/esa/Roth17,DBLP:conf/focs/Roth0W20,DBLP:conf/icalp/Roth0W21,DBLP:conf/iwpec/RothS18,DBLP:conf/mfcs/DorflerRSW19,DBLP:conf/mfcs/PeyerimhoffR0SV21}.
Can these graphs also be constructed in polynomial time with these
techniques? Moreover, when these techniques \emph{do} identify explicit
large-treewidth graphs $T\inhom p$ (e.g., complete graphs), can we
also find graphs $T'\inhom p$ of bounded degree that make Theorem~\ref{thm: poly-mon}
applicable?

Other follow-up works include lower bounds under the exponential-time
hypothesis $\sharpETH$, which would require a ``macroscopic'' version
of a celebrated lower bound for colorful homomorphisms by Marx~\cite{Marx2010}.
Under certain circumstances, it may also be possible to remove cardinality
filters from the main proof. This would yield reductions that require
only two oracle calls, which may render them applicable for $\mathsf{C_{=}P}$-completeness
results~\cite{Curticapean2016a,Hemaspaandra2002}. Moreover, removing
cardinality filters may also make the techniques applicable to modular
counting problems: In their current form, cardinality filters introduce
uncontrollable divisions through their coefficients, while the CFI-based
filters for surjectively $S$-colored graphs only require divisions
by powers of $2$.

\bibliographystyle{plain}
\bibliography{desc_bib}

\end{document}